\documentstyle[preprint,aps]{revtex}
\begin{document}
\draft
\title{New pseudoclassical model for Weyl particles }
\author{D. M. Gitman, A. E. Gon\c {c}alves and I. V. Tyutin
\footnote{Permanent Address: P. N. Lebedev Institute, 53
Leninsky Pr., Moscow 117924, Russia}}
\address{Instituto de F\'{\i}sica, Universidade de S\~ao Paulo
P.O. Box 20516, 01452-990 S\~ao Paulo,\\ S\~ao Paulo, Brazil}
\date{\today}
\maketitle
\begin{abstract}
A new pseudoclassical model to describe Weyl particles is proposed.
Different ways of its quantization are presented. They all lead to
the theory of Weyl particle; namely, the massless Dirac equation and
the Weyl condition are reproduced. In contrast with models discussed
previously, this one admits both the  Dirac quantization
and quasicanonical quantization, with the
fixation of almost all gauge freedom on the classical level.
\end{abstract}

In recent years numerous classical models of relativistic particles
and superparticles have been discussed intensively in different
contexts. First, the interest in such models was initiated by the
close relationship with problems in string theory and gravity, but
now it is clear that it is an important problem itself whether there
exist classical model for any relativistic particle whose
quantization reproduces, in a sense, the corresponding field theory, or
one particle sector in the corresponding quantum field theory. In
this paper we propose a new pseudoclassical model whose
quantization reproduces the quantum theory of Weyl particles. The
history of the question is the following. As is known, first, a
pseudoclassical action of spin$-\frac{1}{2}$ relativistic particle was
proposed by Berezin and Marinov \cite{BM} and after that was
discussed and investigated in many papers \cite{Casa,BCL,BDZVH,BVH,BSSW}.
The action has the form
\begin{equation}\label{Berezin}
S = \int_0^1\left[-\frac{1}{2e}\left(\dot{x}^\mu-i\psi^\mu\chi
\right)^2-\frac{e}{2}m^2-im\psi^5\chi - i\left(\psi_\mu\dot{\psi}^
\mu-\psi_5\dot{\psi}_5\right)
\right]d\tau\;,
\end{equation}
\noindent where $x^\mu,\;e$ are even and $\psi^\mu,\;\chi$ are odd
variables, dependent on a parameter $\tau \in [0,1],\;\mu =
\overline{0,3},\; \eta_{\mu\nu} = {\rm diag }(1-1-1-1).$
Because of the reparametrization invariance of the action, the
Hamiltonian of the model is equal to zero on the constraint surface.
In Refs. \cite{BL,GNR,HA,HA1}, devoted to the quantization of
the model, the authors tried to avoid this difficulty, using the so called
Dirac method of quantization of theories with first-class constraints
\cite{Dirac}. In this method one treats the  first-class
constraints in the sense of restrictions on state vectors.
Unfortunately, in the general case, this scheme of quantization
creates many questions, e.g. with Hilbert space
construction, the nature of the Schr\"{o}dinger equation and so on. A
consistent, but more complicated technical way is to work in
the physical sector, namely, first, on the classical level, to
impose gauge conditions to all first class-constraints to reduce the
theory to one with second-class constraints only, and then quantize
by means of the Dirac brackets (we will call such a method
canonical quantization). First canonical quantization for a
relativistic spin$-\frac{1}{2}$ particle was done in \cite{DI}.
The quantum mechanics constructed there admits the limit $m = 0$. As a
result one gets the quantum theory of massless particles, which is
described by the Dirac equation with $m = 0$, but without any additional
restrictions on the four-spinor $\Psi (x)$( see for example \cite{GG,DIE}):
\begin{equation}\label{Dm0}
i\partial_\mu\gamma^\mu\Psi (x) = 0\;,
\;\;\; \left[\gamma^\mu,\;
\gamma^\nu\right]_{+} = 2\eta^{\mu\nu}\;.
\end{equation}
\noindent It turns out that the
variable $\psi^5$ can be omitted from the action (\ref{Berezin})
at $m=0$. The quantization of such a modified action
reproduces the physical sector in the limit $m=0$ of the massive
quantum mechanics. Unfortunately, such a quantum theory describes
massless spin$-\frac{1}{2}$ particle with all the possible values of
helicity (right and left neutrinos). As is known, the right
(left) neutrino is described by a four-spinor, which obeys, in addition to
the Dirac equation (\ref{Dm0}), the Weyl condition as well:
\begin{equation}\label{Wc}
\left(\gamma^5-\alpha\right)\Psi (x) = 0\;,\;\;\;
\alpha = 1\;(-1)\;,\;\;\;
\gamma^5=i\gamma^0\gamma^1\gamma^2\gamma^3\;.
\end{equation}
\noindent There were several attempts to modify the action
(\ref{Berezin}) at $m = 0$ so that in course of quantization one
can get quantum mechanics with wave functions obeying both
Eqs. (\ref{Dm0}) and (\ref{Wc}) at the same time. So, in
Ref. \cite{HA1} the authors proposed the action (we are using our
notations for this action)
\begin{equation}\label{AH}
S = \int_0^1\left[-\frac{1}{2e}\left(\dot{x}^\mu+g_1Q^\mu
-i\psi^\mu\chi
\right)^2- i\psi_\mu\dot{\psi}^\mu+g_2\left(\Lambda-
\frac{\alpha}{2}\right)
\right]d\tau\;,
\end{equation}
\noindent where $g_a\;, a=1,2$ are Lagrange multipliers, $\Lambda =
i\epsilon^{\mu\nu\rho\varsigma}\psi_\mu\psi_\nu\psi_\rho\psi_\varsigma/3,\;
Q^{\mu} = \epsilon^{\mu\nu\rho\varsigma}\psi_\nu\psi_\rho\psi_\varsigma\;.$
Quantization by means of the Dirac method gives both Eqs.
(\ref{Dm0}) and (\ref{Wc}) as restrictions on state vectors.
Namely, the theory has, in particular, second-class constraints
{\hbox {$P_\mu +i\psi_\mu = 0$}}, where $P_\mu$ are
momenta conjugated to $\psi^\mu$, and first-class constraints
{\hbox {$\pi^2 = 0$}}, $\pi_\mu\psi^\mu = 0,\;\pi_\mu Q^\mu = 0,
\;\Lambda - \alpha /2 = 0$, where $\pi_\mu$ are momenta conjugated
to $x^\mu$. Calculating the Dirac brackets with respect to the second
class constraints only, one can find in the course of quantization
the following realization for essential variables $\hat{\pi}_\mu =
-i\partial_\mu\;,\;\;\;\hat{\psi}^\mu = \frac{i}{2}\gamma^\mu\;,$
in the $x$ representation. Applying the first-class constraints
operators to the state vector, according to Dirac, one gets only two
independent equations
\[
\hat{\pi}_\mu\hat{\psi}^\mu  \Psi (x)=0\;,\;\;
\left(\hat{\Lambda}-\frac{\alpha}{2}\right)
\Psi (x)=0\;,
\]
\noindent which are just Eqs. (\ref{Dm0}) and (\ref{Wc}). As
mentioned before, this way of quantization is not well grounded. Moreover,
attempts to quantize this action canonically fail, since as soon as
one chooses any gauge condition linear in $\psi$, the combination
$\Lambda$ vanishes, and only the Dirac equation remains after quantization.
Another possibility was discussed in \cite{BCDL}. They considered the
theory with the action
\begin{equation}\label{AB}
S = \int_0^1\left\{-\frac{1}{2e}\left[\dot{x}^\mu-
i(\psi^\mu-\frac{2i\alpha}{3}\epsilon^{\mu\nu\rho\varsigma}
\psi_\nu\psi_\rho\psi_\varsigma )\chi\right]^2
-i\psi_\mu\dot{\psi}^\mu\right\}d\tau\;.
\end{equation}
\noindent A formal quantization of the theory following the Dirac
method leads to the equation $i\partial_\mu\gamma^\mu\left(
\gamma^5-\alpha\right)\Psi (x) = 0\;$ \noindent for state vectors,
which is not equivalent to both Eqs. (\ref{Dm0}), (\ref{Wc}).
The canonical quantization  gives the Dirac equation (\ref{Dm0}), but
without any additional restrictions for helicity. That is in
agreement with the fact that classically actions (\ref{AB}) and
(\ref{Berezin})  are equivalent at $m=0$ \cite{BCDL}.

In this paper we propose a new pseudoclassical action
to describe the Weyl particle which is a covariant generalization
of an action \cite{DIE1}. It admits both quasicanonical
quantization (we are fixing gauge freedom which corresponds to two
types of gauge transformations of existing three ones, see below)
and the Dirac quantization. Both of them lead to the theory of Weyl
particles. The new action has the form
\begin{equation}\label{AWp}
S = \int_0^1\left[
-\frac{1}{2e}\left(\dot{x}^\mu-i\psi^\mu\chi-
\epsilon^{\mu\nu\rho
\varsigma}b_\nu\psi_\rho\psi_\varsigma+\frac{i\alpha}
{2}b^\mu\right)^2-i\psi_\mu\dot{\psi}^\mu
\right]d\tau\;,
\end{equation}
\noindent where $x^\mu ,\;e\;,\;\psi^\mu$, and $\chi $ have the same
meaning as in (\ref{Berezin}), the variables $b^\mu$ form an
even four-vector, and $\alpha$ is an even constant.

There are three types of gauge transformations under which the action
(\ref{AWp}) is invariant: reparametrization
\begin{equation}\label{repar}
\delta x^{\mu} =  {\dot{x}}^{\mu} \xi\;,\;\;
\delta e = \frac{d}{d\tau}\left(e \xi\right)\;,\;\;
\delta b^\mu = \frac{d}{d\tau}\left(b^\mu\xi\right)\;,\;\;
\delta \psi^\mu = \dot{\psi}^\mu\xi\;,\;\;
\delta\chi = \frac{d}{d\tau}\left(\chi\xi\right)\;,
\end{equation}
\noindent with an even parameter $\xi(\tau)$; supertransformation
\begin{eqnarray}\label{supert}
&~&\delta x^{\mu}  =  i\psi^{\mu}\epsilon\;,\;\;
\delta e =i\chi\epsilon\;,\;\;
\delta b^\mu  = 0\;,\;\;
\delta \psi^{\mu}  =
\frac{1}{2e}z^\mu\epsilon\;,\;\;
\delta \chi= \dot{\epsilon}\;,\;\;\nonumber\\
&~&z^\mu = \dot{x}^\mu-i\psi^\mu\chi-
\epsilon^{\mu\nu\rho\varsigma}b_\nu\psi_\rho\psi_\varsigma
+\frac{i\alpha }{2}b^\mu\;,
\end{eqnarray}
\noindent with an odd parameter $\epsilon (\tau)$; and an additional
[in comparasion with the action (\ref{Berezin})] gauge transformation
\begin{eqnarray}\label{gauge}
&~&\delta x^\mu = \left(\epsilon^{\mu\nu\rho\varsigma}
b_\nu\psi_\rho\psi_\varsigma-\frac{i\alpha}{2}b^\mu\right)
\kappa\;,\;\;
\delta e = 0\;,
\;\;\delta b^\mu = \frac{d}{d\tau}(b^\mu\kappa)\;,\nonumber\\
&~&\delta\psi^\mu =
\frac{i}{e}\epsilon^{\mu\nu\rho\varsigma}b_\nu
z_\rho\psi_\varsigma\kappa\;,\;\;
\delta \chi = 0\;,
\end{eqnarray}
\noindent with an even parameter $\kappa(\tau)$. The equations of
motion have the form
\begin{eqnarray}\label{EM}
&~&\frac{\delta S}{\delta x^{\mu}} =
\frac{d}{d\tau}\left[\frac{1}{e}z_\mu\right] = 0\;,\;\;
\frac{\delta S}{\delta e} =  \frac{1}{2e^2}z^2 = 0\;,\;\;
\frac{\delta S}{\delta b^\mu} = \frac{1}{e}z^\nu\left(
\epsilon_{\nu\mu\rho\varsigma}\psi^\rho\psi^\varsigma
-\frac{i\alpha}{2}g_{\mu\nu}\right) = 0\;,\nonumber\\
&~&\frac{\delta_r S}{\delta\chi} = \frac{i}{e}z_\mu\psi^\mu = 0\;,\;\;
\frac{\delta_{r}S}{\delta \psi^\mu} =  2i\dot{\psi}_{\mu}-
\frac{1}{e}z^\rho\left(ig_{\mu\rho}\chi+2\epsilon_
{\mu\rho\nu\varsigma}b^\nu\psi^\varsigma\right) = 0\;.
\end{eqnarray}

Going over to the Hamiltonian formulation, we introduce the
canonical momenta
\begin{eqnarray} \label{CM}
&~&\pi_{\mu}~ =\frac{\partial L}{\partial \dot{x}^{\mu}} =
-\frac{1}{e}z_\mu\;,\;\;\;
P_{e}~ = \frac{\partial L}{\partial \dot{e}} = 0\;,\nonumber\\
&~&P_{\chi} = \frac{\partial_{r}L}{\partial \dot{\chi}}
= 0\;,\;\;\;P_\mu = \frac{\partial_{r}L}{\partial \dot{\psi}^\mu} =
-i\psi_\mu\;,\;\;\;
P_{b_\mu}
 = \frac{\partial L}{\partial \dot{b^\mu}} =0\;.
\end{eqnarray}
\noindent It follows from (\ref{CM}) that there exist primary constraints
\begin{equation}
\Phi^{(1)}_{1} = P_e\; ,\;\;
\Phi^{(1)}_{2} = P_{\chi}\;,\;\;
\Phi^{(1)}_{3 \mu} = P_\mu+i\psi_\mu\;,\;\;
\Phi^{(1)}_{4 \mu} = P_{b_\mu}\; .
\end{equation}
\noindent We construct the total Hamiltonian $H^{(1)}$ according
to the standard procedure (we are using the notation of Ref.
\cite{DI}), $H^{(1)} = H+\lambda_{A}\Phi_{A}^{(1)}\;,$ and get, for the $H$,
\begin{equation}\label{H}
H = -\frac{e}{2}\pi^2+i\pi_\mu\psi^\mu\chi-\left(
\epsilon_{\nu\mu\rho\varsigma}\pi^\mu \psi^\rho\psi^\varsigma+
\frac{i\alpha}{2}\pi_\nu\right) b^\nu\;.
\end{equation}
\noindent From the conditions of the conservation of the primary
constraints in time $\tau,\;\dot{\Phi}^{(1)}=\left\{ \Phi^{(1)},
H^{(1)}\right\}=0$, we find secondary constraints
\begin{equation}\label{SC}
\Phi^{(2)}_1 = \pi^2\;,\;\;
\Phi^{(2)}_2 = \pi_\mu\psi^\mu\;,\;\;
\Phi^{(2)}_{3\mu} = T_\mu =
\epsilon_{\mu\nu\rho\varsigma}\pi^\nu\psi^\rho
\psi^\varsigma+i\frac{\alpha}{2}\pi_\mu\;,\;\;
\end{equation}
\noindent and determine $\lambda$, which correspond to the primary
constraint $\Phi^{(1)}_3$. Thus, the Hamiltonian $H$ appears to be
proportional to the constraints, as one could expect in the case
of a reparametrization invariant theory.
No more secondary constraints arise from the Dirac
procedure, and the Lagrangian multipliers, corresponding to the
primary constraints $\Phi^{(1)}_1\;,\Phi^{(1)}_2\;,\Phi^{(1)}_4$,
remain undetermined. One can go over from the initial set of
constraints ($\Phi^{(1)}\;,\Phi^{(2)}$) to the equivalent one
($\Phi^{(1)}\;,\widetilde{\Phi}^{(2)}$), where
$\widetilde{\Phi}^{(2)} = \Phi^{(2)}\left|_{\psi\rightarrow
\widetilde{\psi} = \psi+\frac{i}{2}\Phi_3^{(1)}\;.}
\right.$
\noindent The new set of constraints can be explicitly divided in
a set of the first-class constraints, which is $(\Phi^{(1)}_{1,2},
\Phi^{(1)}_4, \widetilde{\Phi}^{(2)})$ and in a set of second-class
constraints, $\Phi^{(1)}_3$.
Thus, we are dealing with a theory with first-class constraints.
Our goal is to quantize this theory. First,
we will try to impose as much as possible supplementary gauge
conditions to perform canonical quantization. It turns out to be possible
to impose supplementary gauge conditions to all the first-class
constraints, excluding the constraint
${\widetilde{\Phi}}^{(2)}_3$, that corresponds to a fixation of
gauge freedom which corresponds to the two type gauge transformation
(\ref{repar}) and (\ref{supert}). As a result we will have only a set
of first-class constraints, which is a reduction of ${\Phi}^{(2)}_3$
to the rest of constraints. These  constraints we suppose to use
to specify the physical states. All other constraints will be of
secondclass and will be used to form Dirac brackets for canonical
quantization.
Thus, let us impose the gauge conditions $\Phi^G=0$, where
\begin{equation}\label{GC}
\Phi^G_{1} = e+\zeta \pi_{0}^{-1}\;,\;\;
\Phi^G_2 = \chi\;,\;\;
\Phi^G_{3\mu} = b_\mu\;,\;\;
\Phi^G_4 = x_{0}-\zeta \tau\;,\;\;
\Phi^G_5 = \psi^0\;,\;\;
\end{equation}
\noindent where $\zeta = -{\rm sgn}\pi_{0}$.
(The gauge $x_0-\zeta\tau = 0$ was first proposed in Refs.
\cite{DI} as a conjugated gauge condition to the constraint
$\pi^2=m^2$ in the case of scalar and spinning particles. In
contrast with the gauge $x_0 = \tau$, which together with
the continuous reparametrization symmetry breaks the time
reflection symmetry, and therefore fixes the variables $\zeta$,
the former gauge breaks only the continuous symmetry, so that
the variable $\zeta$ remains in the theory to describe states of
particles $\zeta = +1$ and states of antiparticles $\zeta=-1$.
Namely this circumstance allowed one to get Klein-Gordon and
Dirac equations as Schr\"{o}dinger ones in the
course of the canonical quantization.) The requirement
of consistency of the constraint
$\Phi^G,\;\dot{\Phi}^G = 0$
leads to the determination of the Lagrangian multipliers,
which correspond to the primary constraints $\Phi^{(1)}_1$,
$\Phi^{(1)}_{2}$ and $\Phi^{(1)}_4$.

To go over to a time-independent set of constraints we introduce
the variable ${x}_{0}',\;x_{0}' = x_{0}-\zeta\tau$,
instead of $x_{0}$, without changing the rest of the variables.
That is a canonical transformation in the space of all variables
with the generating function {\mbox{$W = x_{0}\pi_{0}'+\tau
\left|\pi_{0}'\right|+W_{0}$}}, where $W_0$ is the generating function
of the identity transformation with respect to all variables except
$x_0$ and $\pi_0$. The transformed Hamiltonian ${H^{(1)}}'$ is of the form
${H^{(1)}}' = H^{(1)} + \partial W/\partial
\tau = H + \left\{\Phi\right\}\; ,$
where $\left\{\Phi\right\}$ are terms proportional to
the constraints and $H$ is the physical Hamiltonian,
\begin{equation}\label{PH}
H =\omega = \left|\mbox{\boldmath $ \pi$}\right|\; ,\;\;
\mbox{\boldmath $\pi  $} = \left(\pi_k\right)\;,\;\;k=1\;,2\;,3\;.
\end{equation}

One can present all the constraints of the theory (including
the gauge conditions), after the canonical transformations, in the
following equivalent form: $K=0,\;\phi = 0,\;T=0,$
\[
K = \left(
e-\omega^{-1}\;,\;\;P_e\;,\;\;\chi\;,\;\;P_{\chi}\;,\;\;
b^\mu\;,\;\;P_{b_\mu}\;,\;\;{x'}_0\;,\;\;|\pi_0|-\omega\;,\;\;
\psi^0\;,\;\;P_0
\right)\;,
\]
\begin{equation}\label{T}
{\phi} = \left(\psi^{{}^\parallel}\;,\;P_k+i\psi_k\right)\;,\;\;
T_0 = \epsilon_{jkl}\pi_j\psi^{k\perp}\psi^{l\perp}-
\frac{i\alpha}{2}\zeta\omega\;,\;\;
T_k = -\zeta\omega\epsilon_{kjl}\psi^{j \perp}\psi^{l \perp}
+\frac{i\alpha}{2}\pi_k\;.
\end{equation}
\noindent [We are using the notations
$a^{k \perp}=\Pi^k_i({\mbox{\boldmath $\pi $}})a^i,\;
a^{{}^\parallel}=\pi_ka^k,\;
\Pi^k_i(\mbox{\boldmath $ \pi $})=\delta^k_i-\omega^{-2}\pi_k\pi_i $].
Both sets of constraints $K$ and $\phi$ are of
second-class, only $T$ is a first-class constraint. The set $K$ has the so
called special form \cite{DI}. In this case if we eliminate the variables
$e,\;P_e,\;\chi,\;P_\chi,\;b^\mu,\;P_{b_\mu},\;{x'}_0,\;|\pi_0|,\;
\psi^0$, and $P_0$ using the above constraints,
the Dirac brackets for the rest of variables with respect to all
second-class constraints $\left(K,\;\phi\right)$ reduce to ones with
respect to the constraints $\phi$ only. Thus, one can only consider
the variables $x^i,\;\pi_i,\;\zeta,\;
\psi^k,\;P_k$, and two sets of constraints, second-class
one $\phi$ and first-class one $T$.
Nonzero Dirac brackets between all the variables have the form
\begin{eqnarray}\label{NZDB}
&~&\left\{x^k, \pi_j\right\}_{D(\phi)} = \delta_j^k\;,\;\;\;\;
\left\{x^k,x^j\right\}_{D(\phi)} =
\frac{i}{\omega^2}\left[\psi^{k\perp},
\psi^{j\perp}\right]_{-}
\;,\nonumber\\
&~&\left\{x^k,\psi^{j\perp}\right\}_{D(\phi)} =
-\frac{\psi^{k\perp}\pi_j}{\omega^2}\;,\;\;\;\;
\left\{\psi^{k\perp},\psi^{j\perp}\right\}_{D(\phi)} =
-\frac{i}{2}\Pi^k_j\left(\mbox{\boldmath $\pi $} \right)\;.
\end{eqnarray}
\noindent Thus, on this stage we have a theory with only
first-class constraints $T$. These constraints
are quadratic in the fermionic variables. On the one hand, that
circumstance makes it difficult to impose a conjugated gauge
condition, on the other hand, imposing these constraints on
states vectors does not create problems with Hilbert space
construction since the corresponding operators of the constraints
have  discrete spectra.
Thus, we suppose to treat only the constraints $T$
in sense of the Dirac method. Namely, commutation relations between
the operators $\hat{x}^k,\;\hat{\pi}_k,\;
\hat{\psi}^k$, which are related to the corresponding classical
variables, we calculate by means of Dirac brackets (\ref{NZDB}), so
that the nonzero commutators are
\begin{eqnarray}\label{NZC}
&~&\left[\hat{x}^k, \hat{\pi}_j\right]_{-} = i\delta_j^k\;,\;\;
\left[\hat{x}^k,\hat{x}^j\right]_{-} =-\frac{1}{\hat{\omega}^2}
\left[\hat{\psi}^{k\perp },\hat{\psi}^{j\perp }\right]_{-}
\;,\nonumber\\
&~&\left[\hat{x}^k,\hat{\psi}^{j\perp }\right]_{-} =
-i\frac{\hat{\psi}^{k\perp }\hat{\pi}_j}{\hat{\omega}^2}\;,\;\;
\left[\hat{\psi}^{k\perp },\hat{\psi}^{j\perp }\right]_{+} =
\frac{1}{2}\Pi^k_j\left(\hat{\mbox{\boldmath $\pi $}}\right)\;.
\end{eqnarray}
\noindent We assume also the operator $\hat{\zeta}$ to have the
eigenvalues $\zeta = \pm 1$ by analogy with the classical theory,
so that $\hat{\zeta}^2  = 1.$
One can construct the realization of the algebra (\ref{NZC}),
above mentioned operator equation for $\hat{\zeta}$ and
equations of constraints $\hat{\psi}^{{}^\parallel} =0,\;
\hat{P}_k + i\hat{\psi}_k = 0$
in a Hilbert space ${\cal R}$, whose elements ${\bf f}\;\in\;
{\cal R}$ are four-component columns depending on ${\bf x}$,
\[
{\bf f} =
\left(
\begin{array}{c}
f_1({\bf x})\\
f_2({\bf x})
\end{array}
\right),
\]
\noindent so that $f_1({\bf x})$ and $f_2({\bf x})$ are two-
component columns.
Such a realization can be found in a similar way to one
used for a spinning particle with $m  \neq 0$ \cite{DI}.
It has the form
\begin{eqnarray}\label{OPER}
&~&\hat{x}^k=x^k{\bf I}+\frac{1}{2\hat{\omega}^2}\epsilon^{kjl}
\hat{\pi}_j\Sigma^l\;,\;\;
\hat{\pi}_k = \hat{p}^k{\bf I} \;,\;\;
\hat{\psi}^{k\perp }  = \frac{1}{2}\Pi^k_j\left(\hat{
\mbox{\boldmath $ \pi $}}\right)
\Sigma^j\;,\;\;
\hat{\zeta}  = \gamma^0 =
\left(\begin{array}{cc}
I & 0\\
0 & -I
\end{array}
\right)\;,
\end{eqnarray}
\noindent where $\hat{p}^k = -i\partial_k\;;\; I$ and
${\bf I}$ are $2\times 2$ and $4\times 4$ unit matrices;
$\mbox{\boldmath $\Sigma $}= {\rm
diag} \left(\mbox{\boldmath $\sigma,\;\sigma $}\right)$,
$\sigma^k$ are Pauli
matrices and $\gamma^0$ is the zeroth $\gamma $ matrix.
The operators $\hat{T}$, which correspond to the first-class
constraints (\ref{T}), have the following form in this realization
\begin{equation}\label{TWc}
\hat{T}_0=\frac{i}{2}\left(
\mbox{\boldmath $\Sigma $}\hat{{\bf p}}
-\alpha\gamma^0\hat{\omega}\right)\;,\;\;
\hat{T}_k=-\hat{p}^k\frac{i}{2}\frac{\gamma^0}{\hat{\omega}}\left(
\mbox{\boldmath $\Sigma $}\hat{{\bf p}}-\alpha\gamma^0
\hat{\omega}\right) \;.
\end{equation}
\noindent Physical state vectors have to obey the conditions
\begin{equation}\label{CSV}
\hat{T}_\mu {\bf f} = 0\;.
\end{equation}
\noindent Their evolution in ``time'' $\tau$ is defined by
the Schr\"{o}dinger  equation
$\left(i\partial/\partial\tau-\hat{H}\right){\bf f}=0$,
which written in terms of the physical time $x^0 =
\zeta\tau$ (see \cite{DI}), has the form
\begin{equation}\label{SE}
\left(i\partial_0-\gamma^0\hat{\omega}\right){\bf f}(x)
= 0\;,\;\;\partial_0=\frac{\partial}{\partial x^0}\;,\;\;
x=(x^0,{\bf x})\;.
\end{equation}

To find a connection of the quantum mechanics constructed with the
theory of Weyl particle let us do the unitary Foldy-Wouthuysen
transformation \cite{FW}, adapted to the case $m=0$:
\begin{equation}\label{FW}
{\bf  f}(x) = {\cal U}\Psi (x)\; ,\;\;\;\;
{\cal U}=\frac{\hat{\omega}+\mbox{\boldmath $\gamma $}{\bf\hat p}}
{\hat{\omega}\sqrt{2}}\; ,\;\;\;{\cal U}^{\dag}{\cal U}=1\;.
\end{equation}
\noindent By straightforward calculations we get
\begin{eqnarray}\label{T0Tj}
&&\check {T} =
{\cal U}^{\dag}\hat{T}_0{\cal U}=\frac{i}{2}\left(\gamma^5-\alpha\right)
\gamma^0\gamma^k\partial_k\;,\nonumber\\
&&\check {T}_j =
{\cal U}^{\dag}\hat{T}_k{\cal U}=-\frac{i}{2}\left(\gamma^5-\alpha\right)
{\partial}_k\;,\nonumber\\
&&{\cal U}^{\dag}\left(i\partial_0-\gamma^0\hat{\omega}\right){\cal U}=
i\gamma^0\gamma^\mu\partial_\mu\;.
\end{eqnarray}
\noindent Thus, after the transformation (\ref{FW})
we get the Dirac equation (\ref{Dm0}) as a consequence of the
Schr\"{o}dinger equation (\ref{SE}). Conditions $\check {T}_\mu\Psi(x)
=0$, which are consequences of (\ref{CSV}), can be rewritten in
the following form for the solutions of the Dirac
equation
\begin{equation}\label{NWC}
\partial_\mu\left(\gamma^5-\alpha\right)\Psi (x)=0\;.
\end{equation}
\noindent Let us chose $\alpha = 1$. Then it follows from
(\ref{NWC}) that
\begin{equation}\label{WF}
\Psi (x)=
\left(
\begin{array}{c}
u(x)\\C
\end{array}
\right)\;,
\end{equation}
\noindent where $C$ is a constant two-component spinor and $u(x)$
a two-component spinor. To provide a finite norm of the wave function
(\ref{WF}), we have to select $C=0$. Thus, on normalized functions
$\Psi (x)$ the Eq. (\ref{NWC}) with $\alpha = 1$ is equivalent to
the Weyl condition (\ref{Wc}) for right neutrino. We have similar
situation in the case $\alpha = -1$, which corresponds to the left neutrino.
So, the action (\ref{AWp}) with $\alpha = 1$ describes the right neutrino
and with $\alpha = -1$ describes the left neutrino.

One can also verify that formal Dirac's quantization of the action
(\ref{AWp}), without any  gauge fixing, leads to the same result.
In this case we have only one set of second-class
constraints $\Phi^{(1)}_3$, which defines Dirac brackets and
commutation relations. For essential operators and
nonzeroth commutators we have
\[
\left[\hat{x}^\mu,\hat{\pi}_\nu\right]_{-}=i
\left\{x^\mu,\pi_\nu\right\}_{D(\Phi^{(1)}_3)}=
i\delta^\mu_\nu\;,\;\;
\left[\hat{\psi}^\mu,\hat{\psi}^\nu\right]_{+}=
i\left\{\psi^\mu,\psi^\nu\right\}_{D(\Phi^{(1)}_3)}=
-\frac{1}{2}g^{\mu \nu}\;.
\]
\noindent As a realization of these commutation
relations one can select in the form
$\hat{\psi}^\mu = \frac{i}{2}\gamma^\mu\;,\;\;\hat{x}^\mu =
x^\mu {\bf I}\;,\;\;\hat{\pi}_\mu = -i\partial_\mu\;.$
\noindent According to Dirac, the  operators
of all the first-class constraints being
applied to the state vectors define the physical states. Using
the primary first-class constraints $\Phi^{(1)}_{1,2,4}$
in this way, one can see that physical vectors are only functions on $x$.
The operators of the secondary first-class constraints
$\hat{\Phi }^{(2)}$, being applied to the state vectors, give the
equations $\hat{\pi}^2\Psi (x) = 0\;,\;\;
\hat{\pi}_\mu\gamma^\mu\Psi (x) = 0\;,\;\;
\hat{T}_\mu\Psi (x) = 0\;.$
\noindent They are equivalent to two sets of independent equations
$\hat{\pi}_\mu\gamma^\mu\Psi (x)=0\;,\;\;
\hat{\pi}_\mu\left(\gamma^5-\alpha\right)\Psi (x)=0\;,$
\noindent which are just Dirac equation (\ref{Dm0})
and the condition (\ref{NWC}).
Therefore, both ways of quantization for the action (\ref{AWp})
give the same result.

The authors would like to thank Professor J. Frenkel for discussions
and one of them (I.V.T.) is grateful to Departamento
de F\'{i}sica Matem\'{a}tica da
Universidade de S\~{a}o Paulo for the kind hospitality and to FAPESP
for its support. A.E.G. was supported by CAPES.

\end{document}